\documentclass[aps,pre,twocolumn,showpacs]{revtex4}
\usepackage{amssymb}
\usepackage{graphicx}
\usepackage{amsmath}
\begin{document}
\title{L\'{e}vy  flights in confining potentials}

\author{Piotr Garbaczewski}\email{pgar@uni.opole.pl}
\affiliation{Opole University, Institute of Physics, Opole, 45-052,
Poland}
\author{Vladimir Stephanovich}\email{stef@uni.opole.pl}
\affiliation{Opole University, Institute of Mathematics and
Informatics, Opole, 45-052, Poland}

\begin{abstract}
We analyze confining mechanisms for L\'{e}vy flights. When they
evolve in suitable external potentials their variance may exist and
show signatures of a superdiffusive transport. Two classes of
stochastic jump - type processes are considered: those driven by
Langevin equation with L\'{e}vy noise and those, named by us
topological L\'{e}vy processes (occurring in systems with
topological complexity like folded polymers or complex networks and
generically in inhomogeneous media), whose Langevin representation
is unknown and possibly nonexistent. Our major finding is that both
above classes of processes stay in affinity and may share common
stationary (eventually asymptotic) probability density, even if
their detailed dynamical behavior look different. That generalizes
and offers new solutions to a reverse engineering (e.g. targeted
stochasticity) problem due to I. Eliazar and J. Klafter [J. Stat.
Phys. 111, 739, (2003)]: design a L\'{e}vy process whose target pdf
equals a priori preselected one. Our observations extend to a broad
class of L\'{e}vy noise driven processes, like e.g. superdiffusion
on folded polymers, geophysical flows and even climatic changes.
\end{abstract}
\pacs{05.40.Jc, 02.50.Ey, 05.20.-y, 05.10.Gg} \maketitle

\section{Introduction}
The study of random walks in complex structures is a key point to
understanding of properties of many physical and non-physical
systems, ranging from transport in disordered media \cite{vank} to
transfer phenomena in biological cells and various real-world
networks  \cite{dor,alb}. It is well-known that a mean square
displacement of a freely diffusing  particle depends on time
linearly $<X^2(t)>\propto t$. If a diffusion is anomalous, then
$<X^2(t)>\propto t^\gamma$, where $\gamma \neq 1$,
$0<\gamma<\infty$. If $\gamma <1$, the dynamics is called {\em
subdiffusive} otherwise {\em superdiffusive}. A superdiffusive
motion of a particle may be generated by means  of non-Gaussian
jump-type processes.

At this point one often invokes  L\'{e}vy flights. Their free
version  may seem somewhat exotic  since their second moments are
nonexistent. However,  L\'{e}vy flights in confining external
potentials show up less exotic behavior  and do  admit the existence
of first few moments (see, e.g., Ref. \cite{lev1}). Thus they may be
employed to model a superdiffusive transport.

L\'{e}vy flights, being non-Gaussian jump-type processes, quite
apart from serious technical difficulties and a shortage of
analytically tractable examples, occur in many fields of modern
statistical physics and have won major attention in the last two
decades \cite{lev1}-\cite{ditlevsen1}. Most of the current research
is devoted to Langevin equation based derivations, where a
deterministic  force is perturbed by the (white) noise of interest,
\cite{cufaro}-\cite{dubkov}. However, in a number of publications,
another class of  jump-type processes was introduced under the name
of topologically induced super-diffusions,
\cite{brockmann}-\cite{geisel1}. The origin of this name is due to
the fact that such processes occur primarily in the systems with
topological complexity like folded polymers or complex networks. An
observation of \cite{sokolov} was that topological super-diffusion
processes do not portray a situation equivalent to any of standard
fractional Fokker-Planck equations and seem  not to correspond to
any  Langevin equation. On the other hand, in the discussion of
above topological L\'{e}vy processes main emphasis has been put on
their super-diffusive behavior with some neglect of confining
effects and the resultant  emergence of  non-Gibbsian stationary
probability densities, \cite{brockmann}-\cite{geisel1}.

We address the latter issue and set general confinement criteria for
an analytically tractable case of Cauchy noise-driven processes. The
results obtained appear to be more general and not specific to
Cauchy noise. To this end, some ideas   have been adopted from the
general theory of diffusion-type stochastic   processes where an
asymptotic approach towards  equilibrium (stationary probability
density function (pdf)) is one of major topics of interest,
\cite{mackey}.

To handle topological L\'{e}vy processes we use a convenient and
general mathematical tool, named Schr\"{o}dinger (or L\'{e}vy -
Schr\"{o}dinger for non-Gaussian processes) semigroup. This tool
naturally appears if one attempts to transform the evolution
equation for the pdf $\rho$ of a certain stochastic process (e.g.
standard or fractional Fokker- Planck equation), into the time -
dependent Schr\"{o}dinger- type equation (the parabolic one in the
Gaussian context; there is no imaginary unit before time derivative)
$\partial _t\psi={\mathcal H}\psi$. Here, $\mathcal H$ receives a
natural interpretation of a Hamiltonian operator, $-\mathcal H$
stands for a semigroup generator. A proper exploitation of a
semigroup operator $\exp(-t{\mathcal H})$ allows not only to
generate the evolution equation for the pdf (differential or
pseudo-differential in case of non-Gaussian L\'{e}vy noise, see
below) but gives access to hitherto unexploited evolution scenarios
which are not captured by the standard Langevin modeling.

We shall demonstrate that topologically induced processes of Refs.
\cite{brockmann}-\cite{geisel1} form a subclass of its solutions
with a properly tailored dynamical semigroup and its (Feynman-Kac)
potential, \cite{klauder,olk}. That allows to take advantage of the
existing mathematical theory of L\'{e}vy processes and L\'{e}vy -
Schr\"{o}dinger semigroups, \cite{applebaum,sato} and
\cite{klauder,olk,olk1},  where  free L\'{e}vy noise generators are
additively perturbed by suitable confining potentials. The theory
works well for both  Gaussian and non-Gaussian processes.

We note here, that in the Brownian case, the Schr\"{o}dinger problem
incorporates the well known transformation of a Fokker-Planck
equation into a generalized diffusion equation, \cite{risken}, e.g.
a transition to the Hermitian (strictly speaking, self-adjoint)
problem whose eigenfunction expansions yield transition pdfs of the
pertinent process.

In this article, we consider an impact of external confining
potentials upon L\'{e}vy flights. The flights may be influenced
directly or indirectly (here via conservative forces) leading to
inequivalent L\'{e}vy processes. An indirect influence refer to
Langevin modeling, while a direct one refers the L\'{e}vy
semigroups. While making this specific distinction between the two
ways of response of L\'{e}vy noise to external potentials, we
address an issue of an apparent incompatibility between them, raised
earlier \cite{sokolov}. The results obtained set a bridge between
these seemingly different classes and may shed some light on the
emergence of varied types of a superdiffusive dynamics in complex
structures, especially those involving significant spatial
inhomogeneities.

\section{Theoretical framework}

\subsection{Smoluchowski processes and Schr\"{o}dinger
semigroups}

To make paper self-contained, here we recapitulate the main
derivations, which will be necessary for us in subsequent
discussion. We begin with consideration of a one-dimensional (1D)
Smoluchowski diffusion process \cite{risken}, with the Langevin
representation $\dot{x} = b(x,t) +A(t)$, where $\langle A(s)\rangle
=0$, $\langle A(s)A(s')\rangle = 2D \delta (s-s')$. Here, $b(x,t)$
is a forward drift of the process, admitted to be  time - dependent,
unless we ultimately pass to Smoluchowski diffusion processes where
$b(x,t) \equiv b(x)$ for all times.

If an initial pdf $\rho_0(x)$ is given, then the diffusion process
drives it in accordance with the Fokker-Planck equation  $ \partial
_t\rho = D\Delta \rho - \nabla \, ( b \rho )$ (in the 1D case
$\nabla \equiv \partial/\partial x$, $\Delta \equiv
\partial^2/\partial x^2$). We introduce an osmotic velocity
field  $u = D\ln \rho $, together with the current velocity field
$v=b - u$. The latter obeys  the continuity equation $\partial _t
\rho = - \nabla j$, where  $j= v\cdot \rho $  has a standard
interpretation of a probability current. The time-independent drifts
$b(x)$ of the diffusion processes are induced by external
(conservative,  Newtonian) force fields $f= - \nabla V $. One
arrives at Smoluchowski diffusion processes by setting
 \begin{equation}\label{enq1}
b = {\frac{f}{m\beta }} = - {\frac{1}{m\beta }}  \nabla V.
\end{equation}
Here, $m$ is a mass and $\beta$ is a reciprocal relaxation time of a
system. The expression \eqref{enq1} accounts for a fully - fledged
phase - space derivation of the spatial process, in the regime of
large $\beta $. It is taken for granted that the
fluctuation-dissipation balance gives rise to the standard form
$D=k_BT/m\beta $ of the diffusion coefficient $D$ ($T$ stands for a
temperature and $k_B$ is Boltzmann constant).

Let us consider a stationary asymptotic regime, where $j\to j_*=0$.
We denote an (a priori assumed to exist, \cite{mackey}), invariant
pdf $\rho _*= \rho _*(x) $. Since in stationary case  $v=v_*=0$, we
have
\begin{equation} \label{enq2}
b_*=u_* = D \nabla  \ln \rho _* .
\end{equation}
Since $b=f/m\beta $ does  not depend on pdf explicitly, $b=b_*$ and
$\rho _*(x) = (1/Z)\, \exp[ - V(x)/k_BT]$. It is seen that our
outcome has Gibbs-Boltzmann form with $Z$ being a partition
function, $Z =\int \exp(-V/k_BT) \ dx$.

Denoting  $F_*\equiv - k_BT \ln Z$, we have
\begin{equation}\label{enq3}
\rho _*(x) = \exp\left( [F_* - V(x)]/k_BT\right) \equiv \exp [2\Phi
(x)]\, .
\end{equation}
Here, to comply with the notations of Refs. \cite{zambrini,klauder}
and with subsequent discussion of a  topological generalization of
the Brownian motion and then L\'{e}vy flights
\cite{brockmann}-\cite{geisel1}, we have introduced a new potential
function $\Phi $ such that $\rho _*^{1/2} = \exp (\Phi) $ and  $b=
2D \nabla \Phi $.

Following a standard procedure \cite{risken}  we transform the
Fokker-Planck equation into an associated Hermitian problem by means
of redefinition $\rho (x,t) = \theta ^* (x,t) \exp [\Phi (x)]$, that
takes the Fokker-Plack equation into  a parabolic one \cite{risken}
$\partial _t\theta _*= D \Delta \theta _* - {\cal{V}} \theta _*$.
Its potential ${\cal{V}}$  derives from a compatibility condition
${\cal{V}}(x) = (1/2)[b^2/(2D) + \nabla b]$.

Smoluchowski process with a unique asymptotic Gibbsian pdf implies
\begin{equation}\label{g1}
{\cal{V}} = D \, {\frac{\Delta \rho _*^{1/2}}{\rho _*^{1/2}}}.
\end{equation}
This equation is a trivialized version (due to the time-
independence of its solution) of the time adjoint equation $\partial
_t\theta = -D \Delta \theta + {\cal{V}} \theta$, see
Refs.\cite{klauder,olk} setting $\theta=\rho_*^{1/2}$.

Introducing ($1/2mD$ rescaled) Schr\"{o}dinger-type  Hamiltonian
${\mathcal H} = -D\Delta + {\cal{V}}$, one arrives at a dynamical
(Schr\"{o}dinger) semigroup operator $\exp(-t{\mathcal H})$,  with
the dynamical rule $ \theta ^*(t) = [ \exp(-t{\mathcal H})\theta
^*](0)$, taking forward the initial data $\theta ^*(x,0)$.

For completeness of discussion, we note that the time adjoint
equation, if applicable, would come out from the reverse time
evolution taking a given final (terminal) $\theta (x,t_{\rm fin})$
backwards in time to $\theta (x,t_{\rm fin}-t)=[ \exp(-t{\mathcal
H})\theta](t_{\rm fin})$, all motions being confined to an interval
$[0,t_{\rm fin}]$.

\subsection{L\'{e}vy - Schr\"{o}dinger semigroups.}

Before passing to an analysis of L\'{e}vy flights, let us set
general rules of the game with respect to the response to external
potentials, once a free noise is chosen. We recall that a
characteristic function of a random variable $X$  completely
determines a probability distribution of that variable. If this
distribution admits a pdf $\rho(x)$, we can write $<\exp(ipX)> =
\int_R \rho (x) \exp(ipx) dx$ which, for infinitely divisible
probability laws,  gives rise to  the famous L\'{e}vy-Khintchine
formula (see, e.g. \cite{applebaum})
\begin{eqnarray}
&&<\exp(ipX)> =\exp \Biggl\{ i\alpha p - (\sigma ^2/2)p^2  +
\nonumber \\
&&+\int_{-\infty }^{+\infty } \left[\exp(ipy) - 1 -
{\frac{ipy}{1+y^2}}\right] \nu (dy)\Biggr \},\label{g2}
\end{eqnarray}
where $\nu (dy)$ stands for so-called L\'{e}vy measure. By
disregarding the deterministic and jump-type contributions in the
above, we are  left with  $<\exp(ipx)>= \exp(-\sigma ^2 p^2/2)$,
hence   $\rho (x)= (2\pi \sigma ^2)^{-1/2} \exp (-x^2/2\sigma ^2)$.

In terms of the  random variable $X_t = (2D)^{1/2} A_t$ of the
Wiener process,  we have $<\exp(ipX_t)>= \exp(-tDp^2)$. By employing
$p \to \hat{p}=-i \nabla $  we identify the semigroup operator
$\exp(tD\Delta )$,  with  $\Delta = d^2/dx^2$. This involves a
special version  ${\mathcal H}=  D\hat{p}^2 =- D\Delta $ of the
general   Hamiltonian ${\mathcal H}=F(\hat{p})$.

From now  on, we concentrate on the integral part of the
L\'{e}vy-Khintchine formula, which is responsible for arbitrary
stochastic jump features. By disregarding the deterministic and
Brownian motion entries we arrive at:
\begin{equation}
F(p) = - \int_{-\infty }^{+\infty } \left[\exp(ipy) - 1 -
\frac{ipy}{ {1+y^2}}\right] \nu (dy),\label{g3}
\end{equation}
where $\nu (dy)$ stands for the appropriate L\'{e}vy measure. The
corresponding non-Gaussian Markov process is characterized by
$<\exp(ipX_t)>= \exp[-t F(p)]$ and yields an operator $F(\hat{p})=
{\mathcal H}$, with $\hat{p} = - i\nabla $.

For the sake of clarity we restrict further  considerations to
non-Gaussian random variables whose pdf's are centered and
symmetric, e.g. a subclass of stable distributions characterized by
\begin{equation}
F(p) = \lambda   |p|^{\mu } \Rightarrow {\mathcal H} \equiv \lambda
|\Delta |^{\mu /2}.\label{g4}
\end{equation}
Here  $\mu <2$ and $\lambda >0$ stands for the intensity parameter
of the L\'{e}vy  process. The  fractional Hamiltonian   ${\mathcal
H}$, which is a non-local pseudo-differential operator, by
construction is positive and self-adjoint on a properly tailored
domain.  A sufficient and necessary condition for both these
properties to hold true is that the pdf of the  L\'{e}vy process is
symmetric, \cite{applebaum}.

The associated  jump-type dynamics is interpreted in terms of
L\'{e}vy flights. In particular
\begin{equation}\label{g5}
F(p)= \lambda  |p| \to {\mathcal H}= F(\hat{p}) =  \lambda |\nabla |
\equiv \lambda (-\Delta )^{1/2}
\end{equation}
refers to the Cauchy process, see e.g. \cite{klauder,olk,olk1}. The
pseudo - differential Fokker-Planck equation, which  corresponds to
the fractional Hamiltonian  (28) and the fractional semigroup
$\exp(-t\hat{H}_{\mu })=\exp(-\lambda |\Delta |^{\mu /2})$, reads
\begin{equation} \label{g5a}
\partial _t \rho  = -  \lambda |\Delta |^{\mu /2} \rho  \, ,
\end{equation}
to be compared with the conventional heat equation $\partial _t \rho
= D \Delta \rho $.

For a pseudo-differential operator $|\Delta |^{\mu /2}$, the action
on a function from  its domain is  greatly simplified (as compared
to L\'{e}vy-Khintchine formula \eqref{g3}), in view of the
properties of the L\'{e}vy measure $\nu _{\mu }(dx)$. We have
\cite{klauder,sokolov,olk,cufaro,dubkov}:
\begin{equation} \label{lkf}
(|\Delta |^{\mu /2} f)(x) = - \int_{-\infty}^\infty [f(x+y) - f(x) ]
\nu _{\mu }(dy).
\end{equation}
The Cauchy-L\'{e}vy measure, associated with the  Cauchy semigroup
generator $|\Delta |^{1/2}\equiv |\nabla |$, reads
\begin{equation}\label{lkf1}
\nu  _{1/2}(dy) = {\frac{1}{\pi }} {\frac{dy}{y^2}}.
\end{equation}

The substitution $y\to z=x+y$ permits to reduce the Eq. \eqref{lkf}
to the familiar form
\begin{equation}\label{lkf2}
 (|\nabla | f)(x) = -  \frac{1}{\pi}
 \int_{-\infty}^\infty {\frac{f(z)- f(x)}{|z-x|^2}}
 dz,
\end{equation}
where $1/\pi |z-x|^2$ has an interpretation of an intensity with
which jumps of the size $|z-x|$ occur.

\section {Response to external potentials: stationary densities}

\subsection{Langevin modeling}
The pseudo-differential Fokker-Planck equation, which  corresponds
to the fractional Hamiltonian  \eqref{g4} and the fractional
semigroup $\exp(-t{\mathcal H}_{\mu })=\exp(-t\lambda |\Delta |^{\mu
/2})$, has the form \eqref{g5a}
to be compared with the Fokker-Planck equation for freely diffusing
particle  (or above heat transfer equation) $\partial _t \rho = D
\Delta \rho $.

In case of jump-type (L\'{e}vy) processes a response to external
perturbations by conservative force fields appears to be
particularly interesting. On one hand, one encounters  a widely
accepted reasoning (Refs. \cite{fogedby}-\cite{dubkov}) where the
Langevin equation, with additive deterministic and  L\'{e}vy "white
noise" terms,  is found to imply a fractional Fokker-Planck
equation, whose form  faithfully  parallels  the Brownian  version,
e.g. (c.f. Ref. \cite{fogedby}, see also \cite{olk1})
\begin{equation}\label{g7}
\dot{x}= b(x)  + A^{\mu }(t)
\Longrightarrow \partial _t\rho =
-\nabla (b\cdot \rho ) - \lambda |\Delta |^{\mu /2}\rho \, .
\end{equation}
Here we make a remark regarding our notations. In 1D case operator
$\nabla$ means simply differentiation over $x$ (see also above) so
that all quantities like $f$ are scalars. In higher dimensions the
operator $\nabla$, acting on vector quantity ${\vec b}\cdot \rho $
(${\vec b}\equiv -{\vec \nabla }V/m\beta$) should be understood as a
vector divergence, i.e. the term ${\vec \nabla }({\vec b}\cdot
\rho)$ $\equiv$ ${\rm {div}} ({\vec b}\cdot \rho)$. Also, here we
emphasize a difference in sign in the second term of Eq. \eqref{g7}
as compared to that in Eq. (4) of Ref. \cite{fogedby}.  There, the
minus sign is absorbed in the adopted definition of the (Riesz)
fractional derivative. Apart from the formal resemblance of operator
symbols, we do not directly employ fractional derivatives in our
formalism.

\subsection{Topological route}
The other approach to account for external perturbations is that, by
mimicking the above Gaussian strategy, we can directly refer to the
Hamiltonian framework and dynamical semigroups with L\'{e}vy
generators being additively perturbed by a suitable potential. For
example, assuming that the functional form of ${\cal{V}}(x)$
guarantees that ${\mathcal H}_{\mu }  \equiv \lambda |\Delta |^{\mu
/2} + {\cal{V}}$ is self-adjoint and bounded from below, we may pass
to the fractional (non-Gaussian, jump process) analog of the
generalized diffusion equation:
\begin{equation}\label{g8}
\partial _t\theta _* = -  \lambda |\Delta |^{\mu /2} \theta _*
- {\cal{V}} \theta _*.
\end{equation}
The dynamical semigroup reads $\exp(- t{\mathcal H}_{\mu })$ and the
compatibility condition related to Eq. \eqref{g1}, takes the form of
the time-adjoint equation $\partial _t\theta = \lambda |\Delta
|^{\mu /2} \theta + {\cal{V}} \theta$  \cite{gar2}. General theory
\cite{klauder,olk,gar2}  tells us that  $\theta ^*(x,t) \theta (x,t)
= \rho (x,t)$ stands for a pdf of an  affiliated   Markov process
that interpolates between the  boundary data $\rho(x,0)$ and $\rho
(x,t_{\rm fin})$, at times  $t\in [0,t_{\rm fin}]$.

We  consider  time-independent $\theta (x,t) \equiv \theta (x)$  and
hereby mimic the Gaussian  ansatz: $\theta (x)= \exp[\Phi (x)]$ so
that $  \theta ^*(x,t) = \rho (x,t)\, \exp [- \Phi (x)]$. If we set
$\exp[ \Phi (x)] = \rho ^{1/2}_*(x)$, we get  the compatibility
condition (see Eq. \eqref{g1}):
\begin{equation}\label{g10}
{\cal{V}}  =   -\lambda\,
{\frac{|\Delta |^{\mu /2}\,  \rho ^{1/2}_*}{\rho ^{1/2}_*}} \, .
\end{equation}
This identity should be compared with Eq. (8) in Ref. \cite{geisel},
where an analogous   effective potential was deduced in the study of
L\'{e}vy flights in inhomogeneous media.

In view of  the semigroup dynamics, we deduce  a continuity equation
with an explicit fractional input
\begin{equation}\label{g11}
\partial _t \rho  = \theta \partial _t \theta ^*=
- \lambda  ( \exp \Phi ) |\Delta |^{\mu /2}[ \exp(-\Phi ) \rho ] +
{\cal{V}} \cdot \rho .
\end{equation}

Up to cosmetic changes $\Phi \to -V/2k_BT$ (compare with Eq.
\eqref{enq3}), Eq. \eqref{g11} is identical with transport equations
employed in a number of papers. There, the investigated process was
named a topologically induced superdiffusion. Namely, with respect
to explicit form of Eq. \eqref{g10}, the Eq. \eqref{g11} assumes a
familiar form of the transport equation (with respect to $\lambda
=1$ and $\kappa =1/(k_BT)$), see Eq. (6) in Ref. \cite{geisel}, Eq.
(5) in Ref. \cite{geisel1} and Eq. (36) in Ref. \cite{brockmann}.
\begin{widetext}
\begin{equation}
\partial _t \rho =-    \exp(-\kappa V/2)\,  |\Delta |^{\mu /2}
\exp(\kappa V/2 )  \rho  + \rho \exp (\kappa V/2) |\Delta |^{\mu /2}
\exp(-\kappa V/2),\label{fk}
\end{equation}
\end{widetext}
We note a systematic  sign  difference between  our $|\Delta |^{\mu
/2}$ and the corresponding fractional derivative $\Delta ^{\mu /2}$
of Refs. \cite{brockmann,geisel,geisel1}.

\subsection{A discord and the reverse engineering problem}\label{sec:dis}

The puzzling point is that for the L\'{e}vy process in external
force fields, the Langevin approach yields a continuity (e.g.
fractional Fokker-Planck) equation in a very different form
\begin{equation}\label{g12}
\partial _t\rho = -\nabla \left(- {\frac{\nabla V }{m\beta }}\,
\rho  \right) - \lambda |\Delta |^{\mu /2}\rho  \, .
 \end{equation}
The conclusion of Refs. \cite{brockmann}-\cite{geisel1} was that,
while  assuming   $\Phi \sim V$ where $V$ is (up to inessential
factors) the above external force potential,  the two transport
equations \eqref{g11} and \eqref{g12} are plainly  incompatible so
that Eq. \eqref{g11} seems not correspond to any Langevin equation
with L\'{e}vy noise term and $b=-\nabla V/m\beta$ as a deterministic
part and vice versa. This puzzling discrepancy has not been explored
previously in more depth.

The problem we address is:
\begin{description}
  \item[(i)] choose a functional form of $V(x)$ and thus
the drift of the Langevin - type  process;
  \item[(ii)] infer an invariant pdf
$\rho _*$ that is compatible  with the  fractional  Fokker-Planck
equation  \eqref{g12};
  \item[(iii)] given $\rho _*$,  deduce  the Feynman-Kac (e.g. dynamical
semigroup)  potential ${\cal{V}}$ by means of Eq. \eqref{g10};
\item[(iv)] use  ${\cal{V}}$ in \eqref{g11}  and  verify whether  the
"topologically induced dynamics" is at all related to that
associated with \eqref{g12} (and thus  to  the underlying Langevin
equation with L\'{e}vy noise);
\item[(v)] check an asymptotic behavior of $\rho(x,t)$ in both
scenarios \eqref{g11} and \eqref{g12} to find possible differences
in the speed (convergence time rate) with which the common invariant
pdf $\rho _*(x)$ from item (ii) is approached;
\item[(vi)] repeat the procedure in reverse order by starting from
step (iii) and then deduce the drift for the Langevin equation with
L\'{e}vy noise; next  compare the dynamical scenarios \eqref{g11}
and \eqref{g12} for any common initial pdf.
\end{description}

We recall that the the  above  problem is non-existent in the case
of Brownian motion. There, the  Fokker-Planck dynamics  and  the
related  parabolic equations do refer to the same diffusion-type
process.

We shall demonstrate below that both Langevin - driven and semigroup
- driven Cauchy processes, albeit non - coinciding literally, keep
resemblance to each other and may share common for both stationary
pdf. A superdiffusive dynamical behavior is generically expected to
arise and an asymptotic approach towards a stationary pdf is then in
principle possible. This motivates the "targeted stochasticity"
discussion whose original formulation (in terms of the reverse
engineering problem) for Langevin - driven L\'{e}vy systems can be
found in Ref. \cite{klafter}. The original formulation of the
reverse engineering problem reads: given a stationary pdf, can we
tailor a drift function so that the system Langevin dynamics would
admit the predefined as an asymptotic target?

We employ the reverse engineering problem to analyze Cauchy
processes in confining potentials. In the course of the discussion,
we in fact extend its range of applicability (that applies to more
general stable processes as well) and demonstrate that a priori
chosen stationary pdf may serve as a target density for both
Langevin and semigroup - driven Cauchy processes. Even though their
detailed dynamical patterns of behavior are different. In the near -
equilibrium regime this dynamical distinction becomes immaterial.

\section{Cauchy driver}

In view of serious technical difficulties we shall not attempt to
present a fully fledged solution to the above formulated problem for
any symmetric stable jump-type process and any conceivable drift.
Instead, we turn our attention to situations where explicit
functional forms of invariant densities are available. Most of them
were inferred in the problems, related to Cauchy noise, see Refs
\cite{klauder,olk1}, \cite{fogedby}-\cite{dubkov}. In particular,
attention has been paid to confining properties of various drifts
upon the Cauchy noise. On the other hand, L\'{e}vy flights through a
"potential landscape" (topological processes of Refs.
\cite{brockmann}-\cite{geisel1}) were interpreted as (enhanced)
super-diffusions.

\subsection {Ornstein - Uhlenbeck - Cauchy process}
Let us consider the Ornstein - Uhlenbeck - Cauchy (OUC) process,
whose drift is given by $b(x)= - \gamma x$, and an asymptotic
invariant pdf associated with the Cauchy-Fokker-Planck equation
$\partial _t \rho = - \lambda |\nabla | \rho + \nabla [(\gamma
x)\rho ]$ reads
\begin{equation}
\rho _*(x) = \frac{\sigma }{\pi} \frac{1}{\sigma ^2 + x^2}, \ \sigma
= \frac{\lambda}{\gamma},\label{g14a}
\end{equation}
c.f. Eq. (9) in Ref. \cite{olk1}. Here, the modified noise intensity
parameter $\sigma $ is  a ratio of an intensity parameter $\lambda $
of the Cauchy noise and of the friction coefficient  $\gamma$. Note
that a characteristic function of this pdf reads $F(p) = - \sigma
|p|$ and accounts for a non-thermal fluctuation-dissipation balance.

For Cauchy random variable $X_t$  we have $ \langle\exp (ip
X_t)\rangle$ = $\exp( t\lambda |p|)$. The corresponding pdf has the
 form \eqref{g14a} with
$\sigma  \sim t\lambda $, e.g. $\rho (x,t)= \lambda t/\pi [(\lambda
t)^2 + x^2]$. Here, $\sigma $ and $t\lambda $  play a role of
scaling parameters specifying the half-width  of the Cauchy pdf at
its half-maximum. Since $t\lambda $ grows monotonically, the
 free Cauchy noise pdf  flattens and its maximum drops down in time.

Since $\sigma = \lambda / \gamma $,  the confining  drift $-\gamma
x$ may stop the "flattening" of the probability distribution and
stabilize the pdf  at quite arbitrary  shape (with respect to its
maximum and half-width, see above), by manipulating $\gamma $. For
example, $\gamma \gg 1$  implies  a significant shrinking of the
distribution $\rho _*$ as compared to the reference (free noise)
pdf at any time $t \sim 1/\lambda $. In parallel, a
maximum pdf value would increase: $1/\pi \lambda \to \gamma / \pi
\lambda $.

The OUC case refers to Cauchy flights in a confining (harmonic)
potential, but does not imply the confined flight, since the
variance of the asymptotic density diverges. We note that confined
L\'{e}vy flights and specifically confined Cauchy flights, have been
analyzed  earlier in Refs. \cite{chechkin}-\cite{dubkov}.

To deduce the  potential ${\cal{V}}$ for the OUC process with given
invariant pdf $\rho _*$, we need  to evaluate the right-hand-side of
the defining  Eq. \eqref{g10}, with $\mu = 1$. We employ Eq.
\eqref{lkf2},  so arriving at:
\begin{eqnarray}
&&\frac{\pi}{\lambda}\frac{1}{(\sigma ^2 + x^2)^{1/2}}\,
{\cal{V}}(x) =\nonumber
\\
&&=\int_{-\infty}^\infty \left[ {\frac{1}{\sqrt{\sigma ^2 +
(x+y)^2}}} - {\frac{1}{\sqrt{\sigma ^2 + x^2}}} \right]
{\frac{dy}{y^2}}.\label{g15}
 \end{eqnarray}
Because of the  integrand singularity at $y=0$, we  must handle  the
integral in terms of its principal value. Introducing the notation
$a=\sigma ^2 + x^2$, we arrive at, \cite{gradstein}:
\begin{equation}\label{lg}
{\cal{V}}(x) = {\frac{\lambda }{\pi }} \left[ - {\frac{2}{\sqrt{a}}}
+ {\frac{x}{a}}\ln {\frac{\sqrt{a}+x}{\sqrt{a}-x}}\right].
\end{equation}
Here, ${\cal{V}}(x)$ is bounded both from below and
above, with the asymptotics  $(2/|x|) \ln |x|$ at infinities, well
fitting to the general mathematical construction  of (topological)
Cauchy processes in external potentials, see Ref. \cite{olk} for
details. The plot of potential \eqref{lg} is reported in Fig.1.

Accordingly, we  know for sure that there exists  a topological
Cauchy process  with the Feyman - Kac potential ${\cal{V}}$, Eq.
\eqref{lg}, whose invariant density coincides with that for the
Langevin - supported  OUC process.

\begin{figure*}[t]
\begin{center}
\includegraphics [width=0.9\columnwidth]{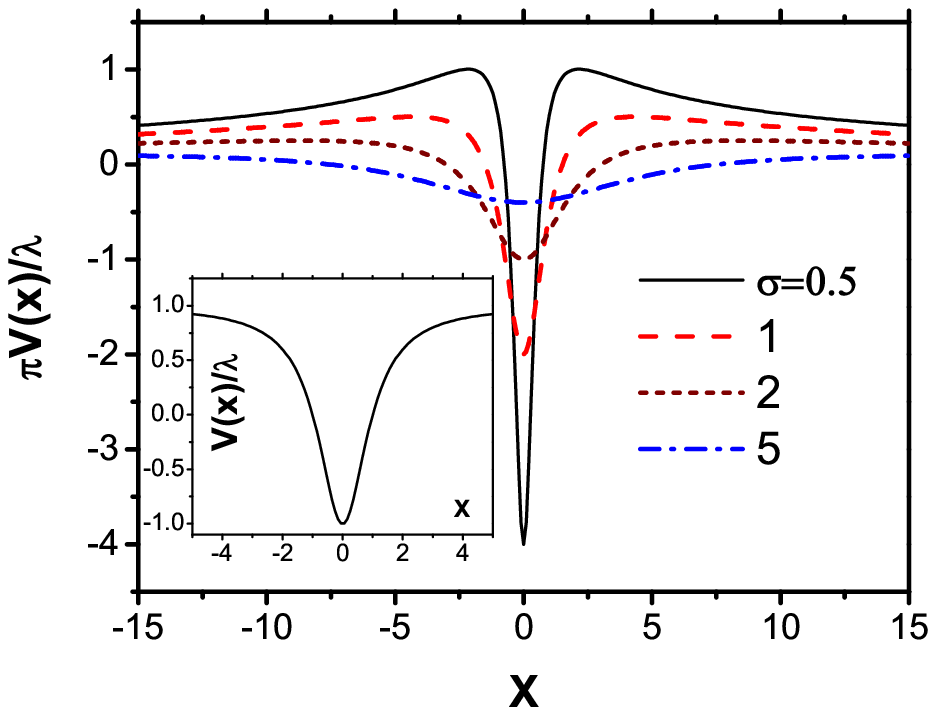}
\includegraphics [width=0.9\columnwidth]{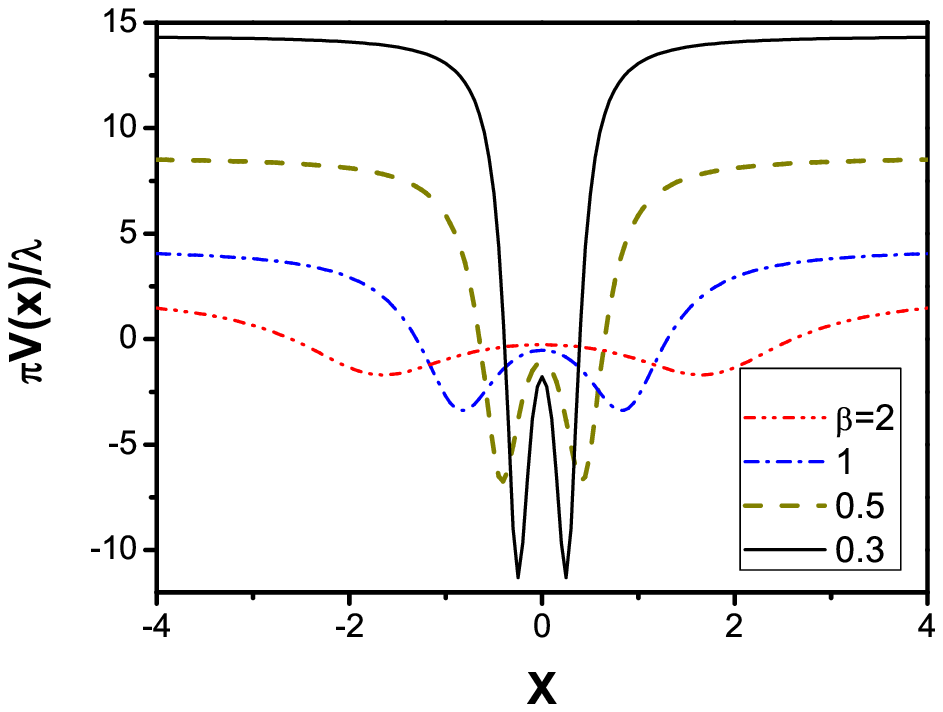}
\end{center}
\caption{The coordinate dependence of potentials ${\cal{V}}(x)$:
\eqref{lg} for different $\sigma$ (main left panel), \eqref{g20}
(inset to left panel) and \eqref{bud2} for different $\beta$ (right
panel).}
\end{figure*}

\subsection{Confined Cauchy processes: Langevin and topological
targeting}

To analyze a time-dependent behavior of both topological and
Langevin - driven process, below we consider specific numerical
example, admitting finite variance $<X^2(t)>$. This time dependent
variance permits to analyze a particular scenario of approaching the
invariant (equilibrium) density in the large time regime. We will
see, that two considered jump - type processes, whose time evolution
is embodied respectively in the fractional Fokker-Planck equation
and in L\'{e}vy-Schr\"{o}dinger semigroup (topological case)
dynamics are definitely alike as they share a common invariant
density. In the  near-equilibrium regime, any dynamical distinction
between these motion scenarios becomes immaterial. However, their
detailed dynamical  behavior far from equilibrium  might be
different and this issue deserves further analytical and numerical
exploration.

To our current knowledge, there is no Langevin - type representation
of a topological process and vice versa, even though an invariant
density is common for both. Nonetheless, we will demonstrate that by
starting from a common initial probability density, the two
(Langevin and dynamical semigroup) motion scenarios end up at a
common invariant density.

Neither OUC process nor its topological counterpart are confined.
For the Cauchy  density, the  second moment is nonexistent. We shall
verify the outcome of the OUC discussion for Cauchy-type processes
whose invariant densities admit the second  moment due to
confinement. Let us consider the quadratic Cauchy pdf:
\begin{equation}\label{cd}
\rho _*(x) = \frac{2}{\pi}\frac{1}{(1+x^2)^2}
\end{equation}

Now, let us proceed in reverse order departing from Eq. \eqref{cd},
so that $(1/\sqrt{2\pi })\rho_*^{1/2} = (1/\pi )/(1+x^2)$ is
actually Cauchy pdf. We consider $f(x) = \rho _*^{1/2} $ as the
initial data for the free Cauchy evolution $\partial _t f = \lambda
|\nabla |f$. This takes $f(x)$ into the form
\begin{equation}\label{g18}
f(x,t)= \sqrt{\frac{2}{\pi }} \, {\frac{1 + \lambda t}{(1 + \lambda
t)^2 + x^2}} .
\end{equation}
Since $\lambda |\nabla |f =  - \lim_{t\to 0} \partial _t f$ we end
up with
\begin{equation}\label{g20}
{\cal{V}} (x) = {\frac{\lim_{t\to 0} \partial _t f}{f}}(x) = \lambda
{\frac{x^2-1}{x^2 +1 }}\, .
\end{equation}
The shape of this potential is shown in Fig.1 (inset to left panel).
A minimum $ -\lambda $ is achieved at $x=0$, ${\cal{V}}=0$ occurs
for $x=\pm 1$, a maximum $+\lambda $ is reached at $x \to \pm \infty
$.

The potential is bounded both from below and above and hence can
trivially be made non-negative (add $\lambda $).  This means that
the potential \eqref{g20} is fully compatible with the general
construction of Ref, \cite{olk}. This topological process is
generated by Cauchy generator plus a potential function, see Ref.
\cite{olk}, is of the jump-type and can be obtained as an $\epsilon
\to 0$ limit of a step process with a minimal step size $\epsilon $.

Note, that in Ref. \cite{olk} no explicit example of the confining
potential ${\cal{V}}$ has been proposed. Eqs. \eqref{lg},
\eqref{g20} provide such examples, which, to our knowledge, have
never been exploited in the literature.

At this point, let us make a guess that the quadratic Cauchy pdf
actually stands for an invariant pdf of  the "normal" Langevin-based
fractional Fokker-Planck equation \eqref{g12} with a drift of the
form \eqref{enq1}.  Accordingly we should  have $\partial _t \rho _*
= 0 = - \nabla (b\, \rho _*) - \gamma |\nabla |\rho _*$ and
therefore the admissible  drift function, if any,  may be  deduced
by means of an indefinite integral:
\begin{equation}\label{g21}
b(x) = -{\frac{\gamma }{\rho _*(x)}} \int (|\nabla |\rho _*)(x)\,
dx.
\end{equation}

For quadratic Cauchy pdf \eqref{cd} the explicit form of $b(x)$
\eqref{g21} reads
\begin{equation}\label{g21a}
b(x) =-\frac{\gamma x}{8}(x^2+3).
\end{equation}
Thus, there exists the Langevin process whose  invariant  pdf is
shared with a corresponding topological process. In the near -
equilibrium regime a dynamical distinction between the pertinent
processes becomes immaterial. In other words, if we wish to deal
with the Langevin process associated with the quadratic Cauchy
density \eqref{cd}, the proper drift form is  given in \eqref{g21a}.

To analyze numerically the above apparent discord between
Langevin-driven and topological processes, we use the invariant pdf
\eqref{cd}, having drift \eqref{g21a} and Feynman - Kac potential
\eqref{g20}. We have chosen this invariant pdf as it has a finite
variance, which permits us to capture the details of near -
equilibrium, initial and intermediate stages of time evolution.

For numerical solution we use simple Euler scheme for time
derivatives and numerical integration (more specifically, we
calculate Cauchy principal value of integrals) on the each Euler
time step for evaluation of fractional derivative $|\nabla|$. The
initial state corresponds to a particle localized at $x=0$,
corresponding to the minima of both potential, derived from the
drift \eqref{g21a} and Feynman - Kac potential \eqref{g20},
$\rho(x,t=0)=\delta(x)$. The solutions $\rho (x,t)$ of the equations
\eqref{g12} (Langevin-type process) and \eqref{fk} (topological
process) are reported in Fig.2 (upper and middle panels
respectively).

It is seen that topological diffusion process needs more time to
achieve the invariant pdf, appears to be slowed down as compared to
the Langevin scenario. This is illustrated in Fig. 2 (lower panel),
where the time evolution of variances for both processes have been
plotted. The time evolution occurs from zero variance of $\delta$
function to asymptotic variance $<X^2(t\to \infty)>=1$ of the pdf
\eqref{cd}. It is seen, that variance for Langevin - type process
achieves the asymptotic value at (dimensionless) time $t\approx
0.5$, while for topological diffusion this time $t\approx 2$. The
shapes of $<X^2(t)>$ for both processes definitely resemble a super
- diffusive motion.

\begin{figure}[!h]
\begin{center}
\includegraphics [width=0.9\columnwidth]{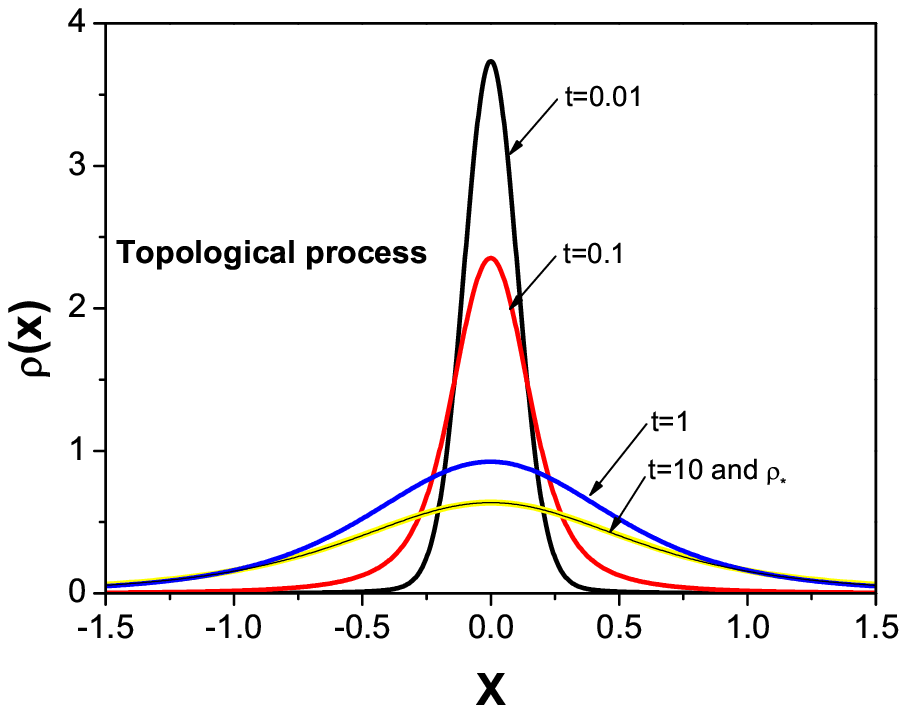}
\includegraphics [width=0.9\columnwidth]{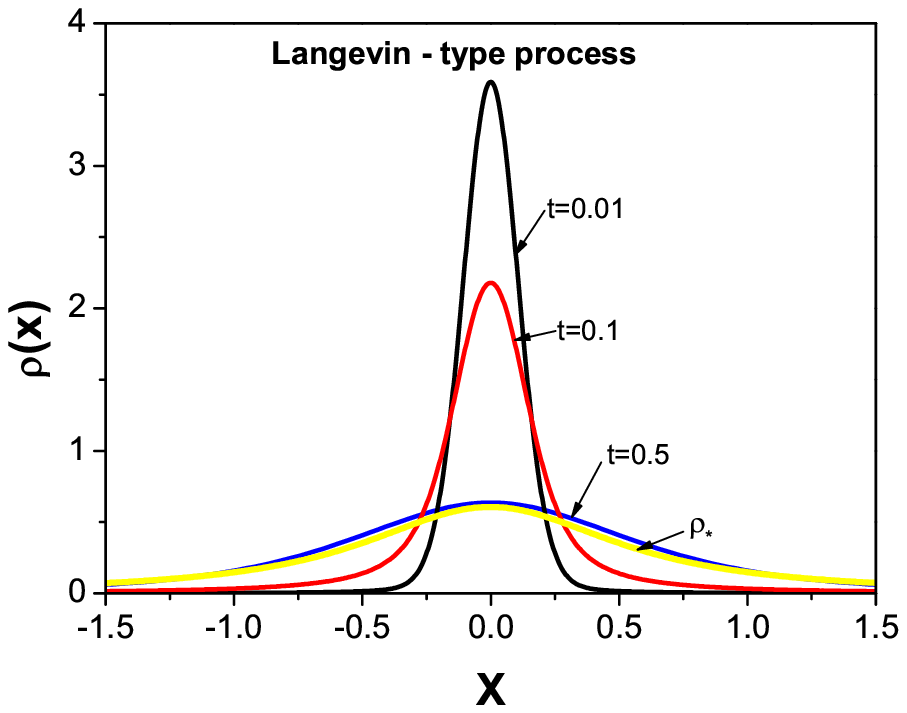}
\includegraphics [width=0.9\columnwidth]{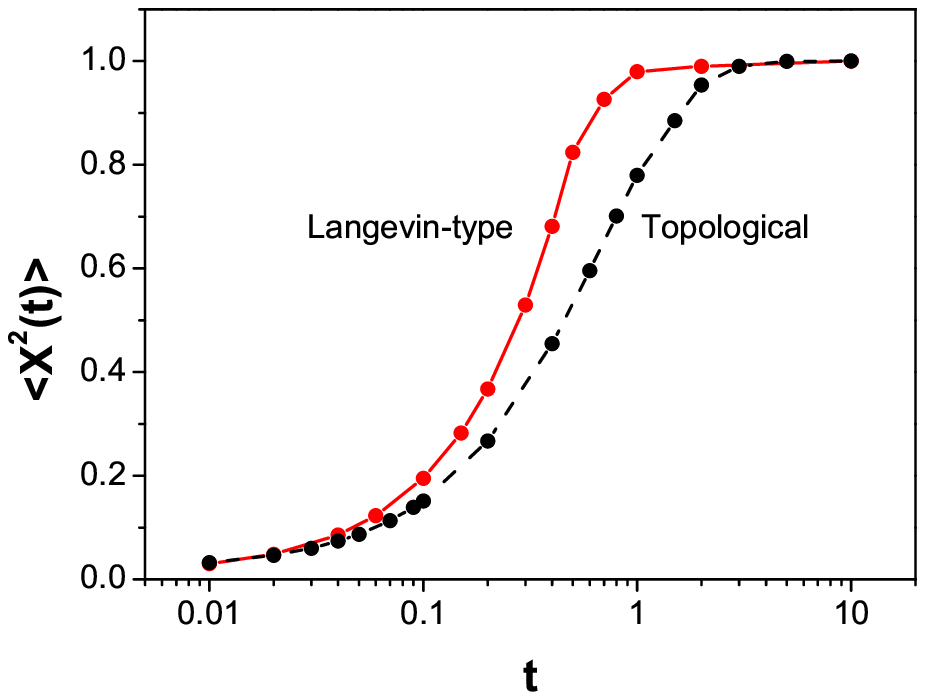}
\end{center}
\caption{Time evolution of pdf's $\rho(x,t)$ for topological (upper
panel, $\lambda=1$) and Langevin-type (middle panel, $\beta=1$,
$m=1$) processes. The common equilibrium pdf $\rho_*$ \eqref{cd} is
also shown. Lower panel reports the time dependent variance $X^2(t)$
for Langevin-type (solid line) and topological (dashed line)
processes. Points correspond to numerical calculation, lines are
guides for eye.}
\end{figure}

\subsection{Confined Cauchy family}
Now we consider a broader class of pdf's related to the Cauchy
noise. Any continuous pdf $\rho $ can be associated with Shannon
entropy $S(\rho ) = -\int \rho \, \ln \rho \, dx$, \cite{kapur}. If
an expectation value $<\ln (1+x^2)>$ is fixed, the maximum entropy
probability function belongs to a one-parameter family
 \begin{equation}\label{ethz}
\rho _* (x)= {\frac{\Gamma (\alpha )}{\sqrt{\pi } \Gamma (\alpha
-1/2))}}\, {\frac{1}{(1+x^2)^{\alpha }}}
 \end{equation}
where $\alpha  >1/2$, \cite{kapur}.

Cauchy distribution is a special case of the above $\rho _*$ that
corresponds to $\alpha =1$. The density \eqref{cd} is the second,
$\alpha =2$, member of the $\alpha $ - integer hierarchy (we assume
that $\sigma =1$).

Our tentative analysis shows that for integer and half - integer
$\alpha$, the invariant pdf \eqref{ethz} admits ${\cal{V}}(x)$,
which fits the restrictions of Corollary 2 in Ref. \cite{olk}. The
question about arbitrary $\alpha$ is still under investigation.

For each specific function ${\cal{V}}(x)$, the resulting Markov
jump-type stochastic process, determined by the Cauchy generator
plus a suitable potential function, appears to be unique. Here we
present only one specific example, namely we consider
\begin{equation}\label{ethz1}
\rho _* (x) = {\frac{16}{5\pi }} \, {\frac{1}{(1+x^2)^4}}.
\end{equation}

Substitution of Eq. \eqref{ethz1} into Eq. \eqref{g10} with respect
to definition \eqref{lkf2} yields \cite{gradstein} the following
expression for the  Feynman-Kac (semigroup) potential
\begin{equation}\label{ethz2}
{\cal{V}}(x)=\frac{\gamma }{2}\frac{x^4 +6x^2 -3}{1+x^2}.
\end{equation}

The potential is bounded from below, its minimum at $x=0$ equals
$-3\gamma /2$. For large values of $|x|$, the potential behaves as
$\sim (\gamma /2) x^2$ i.e. demonstrates a harmonic behavior.

Apart from  the unboundedness  of ${\cal{V}}(x)$ from above, this
potential obeys the minimal requirements of Corollary 2 in Ref.
\cite{olk}: can be made positive (add a suitable constant),
 is  locally bounded (e.g. is bounded  on  each compact set)
and is measurable (e.g. can be approximated with arbitrary precision
by step functions sequences). The  Cauchy generator plus the
potential \eqref{ethz2} determine uniquely an associated Markov
process of the jump-type and its step process approximations.

Having the density \eqref{ethz1}, we can readily address the problem
(vi) of Section \ref{sec:dis}. Namely, inserting Eq. \eqref{ethz1}
to Eq. \eqref{g21}, we obtain
\begin{equation}
b(x) = - {\frac{\gamma x}{16}}\, (5x^6 + 21x^4 + 35 x^2 +35).
\end{equation}
This function shows a linear friction $b\sim - x$ for small $x$ and
a strong taming behavior $b\sim -x^7$ for large $x$.

Let us finally consider a bimodal pdf (see, e.g. Ref. \cite{dubkov})
\begin{equation}\label{ud1}
    \rho_*(x)=\frac{\beta^3}{\pi}\frac{1}{x^4-\beta^2x^2+\beta^4},
\end{equation}
which is a solution of so-called quartic Cauchy oscillator. As a
form of the (confining) potential $V(x)\propto x^4$ is known for
that pdf, we can check the correctness of the procedure \eqref{g21}
of deriving a drift (and hence the potential $V(x)$ in Langevin
scenario) for this pdf. The application of operator \eqref{lkf2} to
function \eqref{ud1} yields
\begin{equation}\label{ud2}
|\nabla |\rho _*(x)=\frac{\pi
x^2}{\beta^3}\frac{x^4+\beta^2x^2-3\beta^4}{(x^4-\beta^2x^2+\beta^4)^2},
\end{equation}
which after integration over $x$ and division over $\rho _*(x)$
\eqref{ud1} yields
\begin{eqnarray}
&& b(x)=-\gamma \frac{x^3}{\beta^3},\label{bud} \\
&& V(x)=-\int b(x)dx= \frac{\gamma}{4\beta^3}x^4,\label{bud1}
\end{eqnarray}
which is exactly the form of the potential for quartic Cauchy
oscillator. The expression \eqref{ud1} can also be used to calculate
the "topological" potential ${\cal V}(x)$
\begin{widetext}
\begin{equation}\label{bud2}
{\cal V}(x)=\frac{\lambda }{\pi }\sqrt {x^4  - \beta ^2 x^2  + \beta
^4 } \int\limits_{ - \infty }^\infty  {\frac{{dy}}{{y^2 }}\left[
{\frac{1}{{\sqrt {(x + y)^4  - \beta ^2 (x + y)^2  + \beta ^4 } }} -
\frac{1}{{\sqrt {x^4  - \beta ^2 x^2  + \beta ^4 } }}} \right]}.
\end{equation}
\end{widetext}
Since an analytic outcome has proved not to be tractable, we have
reiterated to numerics. The result of numerical calculation of the
function \eqref{bud2} is reported in Fig 1 (right panel) for
different $\beta$. It is seen that this potential is also bounded
from below and above, can be made non - negative and have all
properties imposed by Corollary 2 of Ref. \cite{olk}.

\section{Conclusions}
Explicitly solvable  models are scarce in theoretical studies of
L\'{e}vy flights, especially in the presence of external potentials
and/or  external conservative forces. Therefore, our major task was
to find novel analytically tractable examples, that would shed some
light on apparent discrepancies between dynamical patterns of
behavior associated with two different fractional transport
equations that are met in the literature on L\'{e}vy flights.

Although the predominant part of this research is devoted to the
standard Langevin modeling, we have demonstrated that so - called
topological L\'{e}vy processes form a subclass  of solutions to the
Schr\"{o}dinger  boundary data problem. The pertinent dynamical
behavior stems form a suitable L\'{e}vy - Schr\"{o}dinger semigroup.
The crucial role of the involved Feynman - Kac potential has been
identified. We have explicitly derived these potential functions in
a number of cases.

The major gain of above observations is that a mathematical theory
of Ref. \cite{olk} tells one what are the necessary functional
properties of admissible Feynman-Kac potentials. Their proper choice
makes a topological L\'{e}vy process a well behaved mathematical
construction, with a well defined Markovian dynamics and stationary
pdf.

Our focus was upon confinement mechanisms that tame L\'{e}vy flights
to the extent that second moments of their  probability densities
exist. We have shown that the dynamical behavior of both above
classes of processes are close to each other in the near-equilibrium
regime and admit common (for both classes) stationary pdf. This pdf,
in turn, determines a functional form of the aforementioned
(semigroup defining) potential function.

We have generalized the reverse engineering (targeted stochasticity)
problem of Ref. \cite{klafter} beyond the original L\'{e}vy -
Langevin processes setting. We have demonstrated, that within the
targeted stochasticity framework, the concept of L\'{e}vy flights in
confining potentials is not limited to the standard Langevin
scenario. The L\'{e}vy - Schr\"{o}dinger semigroup explicitly
involves confining potentials, but with no obvious link to a
Langevin representation. Our version of the reverse engineering
problem amounts to reconstructing from a given (target) stationary
density the potential functions that either: (i) define the forward
drift of the Langevin process, or (ii) enter the Schr\"{o}dinger -
type Hamiltonian expression in the semigroup dynamics. Both
dynamical scenarios are expected to yield the same asymptotic
outcome i.e. the preselected target pdf.

We note that a departure point for our investigation was a familiar
transformation of the Fokker - Planck operator into its Hermitian
(Schr\"{o}dinger - type) counterpart, undoubtedly valid in the
Gaussian case. The Fokker - Planck and the corresponding parabolic
equation (plus a compatibility condition) essentially describe the
same random dynamics. An analogous transformation is non - existent
for non - Gaussian processes. Two fractional transport equations
discussed in the present paper are inequivalent in the non -
Gaussian case so that the semigroup and the Langevin dynamics with
the L\'{e}vy driver (e.g. noise) refer to different random
processes. The reverse engineering problem allowed us to demonstrate
that those two processes may nevertheless share the same target pdf
and close near equilibrium behavior.

Since the Schr\"{o}dinger boundary data problem allows for a
construction of an  interpolating Markovian processes between any
two a priori prescribed probability densities, it is of interest to
fix an initial pdf and choose an invariant pdf as an asymptotic
(terminal) datum. That is why in the present paper we have given a
detailed comparison of a temporal behavior of the Langevin - based
and topological process, both sharing the same invariant pdf.

\acknowledgements One of us (P. G.) would like to thank Professor J.
Klafter for pointing the reverse engineering idea of Ref.
\cite{klafter} to our attention.

\end{document}